\documentclass{article}

\usepackage[nonatbib, preprint]{neurips_2019}

\usepackage[utf8]{inputenc} 
\usepackage[T1]{fontenc}    
\usepackage{hyperref}       
\usepackage{url}            
\usepackage{booktabs}       
\usepackage{amsfonts}       
\usepackage{nicefrac}       
\usepackage{microtype}      
\usepackage{graphicx}

\title{Deep Adversarial Koopman Model for Reaction-Diffusion Systems}

\author{%
  Kaushik Balakrishnan \\
  Ford Greenfield Labs, Palo Alto, CA\\
  \texttt{kbalak18@ford.com} \\
  \And
  Devesh Upadhyay \\
  Ford Motor Company, Dearborn, MI\\
  \texttt{dupadhya@ford.com} \\
}

\begin{document}

\maketitle

\begin{abstract}
Reaction-diffusion systems are ubiquitous in nature and in engineering applications, and are
often modeled using a non-linear system of governing equations. While robust numerical 
methods exist to solve them, deep learning-based reduced order models (ROMs) are
gaining traction as they use linearized dynamical models to advance the 
solution in time. One such family of algorithms is based on Koopman theory, and this
paper applies this numerical simulation strategy to reaction-diffusion systems. 
Adversarial and gradient losses are introduced, and are found to robustify the predictions. 
The proposed model is extended to handle missing training data as well as recasting
the problem from a control perspective. The efficacy of these developments are demonstrated for two 
different reaction-diffusion problems: (1) the Kuramoto-Sivashinsky equation of chaos and (2) the
Turing instability using the Gray-Scott model.  
\end{abstract}

\section{Introduction}

Reaction-diffusion systems are copious in nature and researchers often represent them as a system of
partial differential equations (PDEs), which are solved using robust numerical methods such as finite difference or
finite volume methods. 
Reaction-Diffusion problems are used to explain patterns observed in the natural world, chaos, combustion, electrochemistry, etc., 
and thus accurate solution procedures are warranted. Many engineering systems in aerospace and automotive propulsion also involve
a reaction-diffusion problem. Often, these reaction-diffusion systems are modeled as a dynamical system
with sophisticated numerical models to advance the solution in time. Computational Fluid Dynamics (CFD) solution procedures 
are accurate and widely used, but are time consuming to solve since the number of degrees of freedom are usually
large. Thus, recasting the problem in a lower-dimensional space and advancing the solution in time in this reduced dimensional space is
widely attempted. To this end, many reduced order models (ROMs) have been developed by the research community with varying 
degrees of mathematical complexity to investigate such systems, and have demonstrated good amount of success \cite{POD1, POD2, DEIM}. 
While these references are based on Proper Orthogonal Decomposition (POD), ROMs based on Koopman models \cite{Koopman1931, Koopman1932} are also
considered where a non-linear model is assumed to have linear dynamics in a Koopman-invariant subspace. Here, numerical methods such as 
dynamic mode decomposition (DMD) \cite{Schmid2010, Kutz2016} are popular. 

In parallel, the Deep Learning community has also invented novel and efficient numerical algorithms for a 
variety of problems in computer vision, robotics, audio synthesis, natural language processing, etc. 
For example, neural networks-based autoencoders have been developed for dimensionality reduction \cite{AE} and are 
solved using the Backpropagation algorithm \cite{Backprop}. 
Neural networks are universal function approximators \cite{Hornik1989, Hornik1991} and so they are fairly robust to 
represent very high-dimensional data. Generative models such as Generative Adversarial Networks (GANs) \cite{GAN} have also been developed by the 
deep learning community and have demonstrated excellent progress for many applications such as computer vision, speech synthesis, and robotics. 
GANs are the new gold standard in deep generative models and many flavors of GANs have been developed in these fields.

In recent years, the ROM and Deep Learning developments have come together to evolve deep learning-based reduced order models for
solving dynamical systems \cite{Takeishi2017, Yeung2017, Lusch2018, Morton2018, Morton2019}. These papers focused on neural network-based 
Koopman models and \cite{Lusch2018, Morton2018} applied it to the von Karman vortex shedding \cite{vonKarman} problem behind a cylinder. 
In \cite{Lusch2018}, an auxiliary neural network was trained to obtain the Koopman eigenfunctions, whereas in \cite{Takeishi2017, Morton2018}
a least squares approach was used. In \cite{Yeung2017}, a global Koopman operator was learned as part of the neural 
network optimization, which however may not be possible in other engineering problems. 
The problems considered in these \cite{Lusch2018, Morton2018} are primarily periodic (or quasi-periodic) and so the model was used to predict future states of the
dynamical system. These methods are powerful, but further studies on applying them to problems that are not periodic is warranted, since 
many real-world engineering and scientific problems many not always be periodic. Furthermore, it is also of interest to investigate if 
such Koopman-based models can be applicable in noisy and/or missing data scenarios.         

In this investigation, we couple a GAN with the autoencoder-based Koopman model of \cite{Lusch2018, Morton2018} and apply it to
two classes of reaction-diffusion systems. First, we investigate a chaotic system in 1D based on the
Kuramoto-Sivashinsky equation \cite{Kuramoto1978, Sivashinsky1977} and demonstrate the robustness of our method. 
By including the GAN loss term in the deep Koopman model at training, the error in the test time predictions are
relatively lower, as will be demonstrated in this paper. 
In the second problem, we consider the Gray-Scott model \cite{GrayScott, Pearson1993} in 2D and obtain good predictions of the Turing
structures. We also consider two alternate settings: (1) a missing data problem where some of the data is not available and the model
is tasked to generate these missing entries; (2) a control problem where we add control inputs to the model in the neural embedding space   
to modify the dynamical evolution of the system. In both these problems, we demonstrate the efficacy of our model. 
While we have considered reaction-diffusion systems only in this study, the proposed model is general and can
be easily extended to other high-dimensional dynamical systems \textit{mutatis mutandis}.


\section{The Koopman operator for dynamical systems}

We consider dynamical systems of the form: 
\begin{equation}
\frac{\partial x}{\partial t} = F\left(x \right ),
\end{equation}
where $x$ is a state vector with $x \in \mathcal{R}^N$ and $F$() is the function that describes the dynamics of the 
system. For instance, $F$ would be the Navier-Stokes equations for fluid dynamic applications, the Schrodinger equation for quantum
mechanics, Maxwell's equations for electrodynamics, etc. While dynamical systems can be either continuous or discrete in time, the 
above form of the equation represents a dynamical system in the continuous setting. This paper will focus only on discrete dynamical systems, and 
in this setting, the dynamical system evolves in time as:
\begin{equation}
x_{t+1} = F\left( x_t \right).
\end{equation}
Most dynamical systems are non-linear and so one cannot obtain a linear system of the form $x_{t+1} = \mathcal{A} x_t$ to evolve in time from $t$ to $t+1$, with 
$\mathcal{A} \in \mathcal{R}^{N\times N}$. For instance, for a linear dynamical system, if we have snapshots in time $x_{1:T+1}$ up to $T+1$, one can construct the 
matrices:
\begin{eqnarray}
X &=& \left[x_1, x_2, \cdots, x_T \right ] \nonumber \\
Y &=& \left[x_2, x_3, \cdots, x_{T+1} \right ] 
\label{eq:XY}
\end{eqnarray} 
and then determine a matrix $\mathcal{A}$ such that $\mathcal{A} = Y X^{\dagger}$ where $X^{\dagger}$ is the Moore-Penrose pseudoinverse of $X$.
This is only valid for a linear dynamical system. However, since most real-world dynamical system are non-linear, this linearized 
treatment will be erroneous. This calls for the development of alternate linearizations of non-linear dynamical systems.

\subsection{Vanilla Koopman dynamical model}

Consider the discrete dynamical system described as $x_{t+1} = F(x_t)$ with $x \in \mathcal{R}^N$. 
In the vanilla Koopman dynamical model \cite{Koopman1931, Koopman1932}, the state vector $x_t$ is mapped on to a Hilbert space 
of possible measurements y = g(x) of the state. In this Koopman invariant subspace, the evolution of the 
system in time is linear, and the infinite-dimensional Koopman operator $\mathcal{K}$ advances the system as:

\begin{equation}
\mathcal{K} g \left( x_t \right) = g \left( F \left( x_t \right) \right) =   g \left( x_{t+1} \right).
\end{equation}
The system is then re-projected back to the state vector space using an inverse function $g^{-1}$ \cite{Proctor2014, Kutz2016, Kaiser2017}:

\begin{equation}
g^{-1} \,\, \left( \mathcal{K} g \left( x_t \right) \right) = x_{t+1}.  
\end{equation}
In practice, however, evaluating such functions $g$ and $g^{-1}$ for many real world systems poses a challenge \cite{Mezic2004, Mezic2013, Lusch2018}.   
Finite-dimensional representations of the Koopman operator can be obtained using Dynamic Mode Decomposition (DMD) \cite{Schmid2010, Kutz2016} for many
problems. In DMD, spatio-temporal coherent structures are identified from a high-dimensional dynamical system. However, since it is based on a 
linearized analysis, it does not generally capture non-linear transients \cite{Lusch2018}.

\subsection{Deep Koopman dynamical model}

To overcome the challenges in evaluating accurate $g$ and $g^{-1}$ functions, recent research works have addressed this with the use
of emerging Deep Learning algorithms combined with large data sets \cite{Takeishi2017, Lusch2018, Morton2018} and have demonstrated
good success for a variety of real world problems. Specifically, an autoencoder \cite{AE} is used to represent the functions $g$ and $g^{-1}$. 
The autoencoder comprises of two neural networks, an
encoder to approximate the function $g$, and a decoder to approximate $g^{-1}$. The use of neural networks 
is a natural choice to represent complex functions as they are universal function approximators \cite{Hornik1989, Hornik1991}. 
In this setting, an input data $x_t$ is mapped to the 
embedding space $z_t$ via the encoder as $z_t = g(x_t)$, which is then mapped through the decoder to obtain 
a reconstruction of the original data point as $\widehat{x_t} = g^{-1}(z_t)$. We represent the dimensionality as
$x \in \mathcal{R}^N$ and $z \in \mathcal{R}^M$, with $M \ll N$.
The autoencoder mappings can be 
mathematically represented as $g : \mathcal{R}^N \rightarrow \mathcal{R}^{M}$ and $g^{-1} : \mathcal{R}^M \rightarrow \mathcal{R}^{N}$.
For evolving the system dynamics in time, the analysis is extended as follows. 

Consider time snapshots of the system $x_{1:T+1}$ and the matrices $X$ and $Y$
introduced in Eqn. (\ref{eq:XY}). These snapshots are fed into the encoder $g$ to obtain latent embeddings $Z = [z_1, z_2, \cdots, z_{T+1}]$, which can
be written in a form similar to Eqn. (\ref{eq:XY}), albeit in the latent embedding space:

\begin{eqnarray}
Z &=& \left[z_1, z_2, \cdots, z_T \right ] \nonumber \\
Z_{+1} &=& \left[z_2, z_3, \cdots, z_{T+1} \right ] 
\label{eq:Z}
\end{eqnarray} 
In \cite{Takeishi2017, Morton2018} least squares fit were undertaken to evaluate an $K$-matrix that can propagate the latent embeddings in time as
$K = Z_{+1} Z^{\dagger}$. Subsequently, $Z$ and the propagated state embeddings advanced in time, $Z_{+1}^{\mathrm{pred}}$, are
fed into a decoder $g^{-1}$ to obtain the reconstructed $\widehat{X}$ and the time advanced solution $\widehat{Y}$. 
The Deep Koopman model is trained to minimize the loss \cite{Takeishi2017, Morton2018}:

\begin{equation}
\mathcal{L} = \parallel X - \widehat{X} \parallel^2 + \parallel Y - \widehat{Y} \parallel^2.    
\end{equation}
Here, the first term in the loss function enforces the autoencoder constraint that the decoded output is approximately close to the input.
The second term in the loss function ensures that the system dynamics' time evolution is captured. 

In \cite{Takeishi2017}, the time advanced latent embedings were evaluated as $Z_{+1}^{\mathrm{pred}} = K \, Z$, i.e., by 
advancing the dynamics by one time step for each snapshot. 
This, however, may not be very accurate at test time when one must predict the state of the system over a long time duration. 
To overcome this problem, a second variant was proposed in \cite{Morton2018} where $Z_{+1}^{\mathrm{pred}}$ was evaluated by applying the 
$K$-matrix recursively from the embedding of the first time step snapshot, i.e., $g(x_1)$.
(Note that \cite{Morton2018} used the symbol $A$ for the matrix, which we rename as $K$ to refer to the Koopman matrix.) 
Specifically, in this second variant \cite{Morton2018}:

\begin{equation}
Z_{+1}^{\mathrm{pred}} = \left[K \, z_1, K^2 \, z_1, \cdots, K^{T} \,z_1 \right].
\label{eqn:rec}
\end{equation}    
(Note that $K^T$ is $K$ raised to the power of $T$ and not the transpose.) The authors of \cite{Morton2018} used time sequences of 32 and demonstrated lower
errors vis-\`a-vis the one time step dynamical advancement version (i.e., \cite{Takeishi2017}) for the 
von Karman vortex shedding problem \cite{vonKarman, Roshko1955}. In a third variant, \cite{Lusch2018} also considered other
losses, including a loss in the embedding space: $\parallel Z_{+1}^{\mathrm{pred}} - Z_{+1} \parallel^2$ with $Z_{+1}^{\mathrm{pred}}$ obtained as shown in
Eqn. (\ref{eqn:rec}), and also $L_{\infty}$ losses to penalize the data point with the largest loss. We will now summarize the Deep Adversarial Koopman model.

\subsection{Deep adversarial Koopman dynamical model}

\subsubsection{GAN objective}
In this study, we propose another variant of the deep Koopman operator. Specifically, we couple a 
Generative Adversarial Networks (GAN) Discriminator \cite{GAN} with the deep Koopman operator
to obtain a Deep Adversarial Koopman Operator. Coupling a GAN Discriminator with an 
autoencoder can improve the quality of samples output from the autoencoder \cite{VAEGAN}. 
The feature representations learned by the GAN discriminator are leveraged to improve the overall quality of the
autoencoder outputs, by including additional loss terms \cite{VAEGAN}, which in the context of reduced order dynamical
systems is novel.
 
In \cite{Lusch2018, Morton2018}, the authors used 
a sequence of snapshots to train the model more accurately, and this approach is used in this study. 
We train the network at each iteration step using a randomly sampled contiguous sequence of length $n_S$, where the 
first entry in the sequence is represented as $x_t$ for this discussion.
Let us consider the sequences: 
\begin{eqnarray}
X &=& \left[x_{t}, x_{t+1}, \cdots, x_{t+n_S-1} \right ] \nonumber \\
X_{+1} &=& \left[x_{t+1}, x_{t+2}, \cdots, x_{t+n_S} \right ] \nonumber \\
X_{+1}^{\mathrm{pred}} &=& \left[x_{t+1}^{\mathrm{pred}}, x_{t+2}^{\mathrm{pred}}, \cdots, x_{t+n_S}^{\mathrm{pred}} \right ].
\label{eq:gan1}
\end{eqnarray}
The deep adversarial Koopman model takes $x_t$ as input and outputs the sequence $X_{+1}^{\mathrm{pred}}$ using the Koopman 
dynamics recursively $n_S$ times. The ground truth values $X_{+1}$ from the data is now used to construct the 
GAN losses. Specifically, we have two concatenated pairs: ($X,X_{+1}$) which we will term as ``real'' and 
($X,X_{+1}^{\mathrm{pred}}$) which we will term as ``fake,'' similar to the parlance used in GAN literature.  
The GAN discriminator takes as input one of these concatenated pair and outputs a single real value $D(\cdot)$, which is
used to construct the GAN objective, following the Wasserstein GAN \cite{WGANGP} approach due to its robustness against mode collapse: 

\begin{equation}
L^{\mathrm{GAN \, objective}} = \mathop{\mathbb{E}}_{x\in (X,X_{+1})} \left[ D(x) \right ] - \mathop{\mathbb{E}}_{\widetilde{x}\in (X,X_{+1}^{\mathrm{pred}})} \left[ D(\widetilde{x}) \right ]. 
\end{equation}

\subsubsection{Loss terms}
The encoder and decoder are represented as $g(\cdot)$ and $g^{-1}(\cdot)$, respectively. 
The loss terms used are mean squared errors (MSE) and the different loss terms we 
consider are: (1) reconstruction loss $L^{\mathrm{recon}}$, (2) prediction loss $L^{\mathrm{pred}}$,
(3) code loss $L^{\mathrm{code}}$, (4) gradient loss $L^{\mathrm{grad}}$, (5) $L_2$ regularization loss $L^{\mathrm{reg}}$,
(6) GAN loss $L^{\mathrm{GAN}}$, and (7) discriminator loss $L^{\mathrm{disc}}$. These different losses are summarized below:

\begin{equation}
L^{\mathrm{recon}} = \parallel x_t - g^{-1} g \left( x_t \right)  \parallel_{\textrm{MSE}}
\end{equation}

\begin{equation}
L^{\mathrm{pred}} = \frac{1}{n_S} \sum_{m=1}^{n_S} \parallel x_{t+m} - g^{-1} \left( K^m g \left( x_t \right) \right) \parallel_{\textrm{MSE}}
\end{equation}

\begin{equation}
L^{\mathrm{code}} = \frac{1}{n_S} \sum_{m=1}^{n_S} \parallel g \left( x_{t+m} \right) - K^m g \left( x_t \right)  \parallel_{\textrm{MSE}}
\end{equation}

\begin{eqnarray}
L^{\mathrm{grad}}_k &=& \frac{1}{n_S} \sum_{m=1}^{n_S} \parallel \nabla_k \left[ x_{t+m} - g^{-1} \left( K^m g \left( x_t \right) \right) \right]  \parallel_{\textrm{MSE}}, \textrm{k=1, 2, 4} \nonumber \\
L^{\mathrm{grad}} &=& \lambda_1 L^{\mathrm{grad}}_1 + \lambda_2 L^{\mathrm{grad}}_2 + \lambda_4 L^{\mathrm{grad}}_4
\end{eqnarray}

\begin{equation}
L^{\mathrm{reg}} = \lambda_{\mathrm{reg}} \sum w_i^2 
\end{equation}

\begin{equation}
L^{\mathrm{GAN}} = \mathop{\mathbb{E}}_{\widetilde{x}\in (X,X_{+1}^{\mathrm{pred}})} \left[ D(\widetilde{x}) \right ]
\end{equation}

\begin{equation}
L^{\mathrm{disc}} = \mathop{\mathbb{E}}_{\widetilde{x}\in (X,X_{+1}^{\mathrm{pred}})} \left[ D(\widetilde{x}) \right ] - \mathop{\mathbb{E}}_{x\in (X,X_{+1})} \left[ D(x) \right ]
\end{equation}

\noindent Note that the losses $L^{\mathrm{recon}}$, $L^{\mathrm{pred}}$,
and $L^{\mathrm{code}}$ were also considered in \cite{Lusch2018}. 
Furthermore, note also that $K^m$ in the equations is a notation that implies that the Koopman model is applied $m$ times, since we use
an auxiliary network \textit{AUX} to obtain $K$, unlike \cite{Morton2018} (more discussion below).   
In addition, we include gradient losses to improve quality \cite{Mathieu2015} of the output
(the terms $\nabla_1$, $\nabla_2$ and $\nabla_4$ refer to the first, second and fourth derivative).

We have a total of four neural networks to train: the encoder, decoder, auxiliary network and the discriminator. 
The neural network architectures are summarized in Appendix A. They are trained using the 
total loss function:

\begin{equation}
L^{\mathrm{total}} = L^{\mathrm{recon}} + L^{\mathrm{pred}} + L^{\mathrm{code}} + \lambda_{\mathrm{grad}} L^{\mathrm{grad}} + \lambda_{\mathrm{reg}} L^{\mathrm{reg}} + \lambda_{\mathrm{GAN}} L^{\mathrm{GAN}}. 
\end{equation}
The $\lambda$ values were experimented to obtain the best results. 
All the results reported in this paper are with $\lambda_{\mathrm{reg}}$ = 10$^{-3}$. When we use the gradient loss, $\lambda_{\mathrm{grad}}$ = 1, otherwise it is set to 0.
Likewise, when we use the GAN loss, $\lambda_{\mathrm{GAN}}$ = 0.01, otherwise it is set to 0. The gradient loss terms are summarized in Appendix B.
The discriminator is trained by minimizing $L^{\mathrm{disc}}$, along with an additional gradient penalty loss term similar to WGAN-GP \cite{WGANGP}. 
In the training of GANs, multiple discriminator update steps are often used per update step of the generator \cite{GAN}. 
We use 4 update steps of the discriminator for each update of the deep adversarial Koopman model.  

\subsubsection{Auxiliary network}
Obtaining the Koopman matrix $K$ for non-linear problems can be very challenging.  
In \cite{Morton2018}, a simple least squares approach was used to obtain $K$.
In \cite{Lusch2018}, an auxiliary neural network was used to obtain the eigenvalues of $K$, 
from which the $K$ matrix was constructed using Jordan block structure. 
The problems under study in this paper do not conform $K$ to any particular structure, and
so we use an auxiliary network to output $K$ at every time step. 
This auxiliary network takes the embedding latent code $z_t$ as input and outputs
$K$, so it is a mapping from $\mathcal{R}^{M} \rightarrow \mathcal{R}^{M \times M}$.
Fully connected layers are used in the auxiliary neural network, and the 
network architecture is presented in Appendix A.

\subsubsection{Residual Koopman}
We also use a residual Koopman model for the dynamics. 
In the standard Koopman time advancement, the model of \cite{Lusch2018, Morton2018} use $z_{t+1} = K z_{t}$ for advancing the dynamics in time.
Our experiments however reveal that a slight modification to this is more robust: we undertake the Koopman dynamics as a residual operation: $\mathcal{K} z_{t} = z_{t+1} = z_{t} + K z_{t}$. 
This simple modification worked well for all the experiments, and so in this setting the Koopman dynamics learns the residual change in the latent 
embedding space from one time step to the next. 
The auxiliary network that is used to learn $K$ converges well with this residual Koopman dynamic model. 
Furthermore, for the recursive multi-step time advancement, we replace Eqn. (\ref{eqn:rec}) with:
\begin{equation}
Z_{+1}^{\mathrm{pred}} = \left[\mathcal{K} \, z_1, \mathcal{K}^2 \, z_1, \cdots, \mathcal{K}^{T} \,z_1 \right],
\end{equation}
where $\mathcal{K}^n$ represents the application of the Koopman opertaor $\mathcal{K}$ $n$ times.


\section{Experiments}

We will now demonstrate the robustness of the Deep Adversarial Koopman model on two reaction-diffusion problems. 
Then, we will show how it can be applied to two variants of the problem: (1) missing data problem and (2) coupling it with control inputs.

\subsection{Test cases}

The focus of this paper is on applying Deep Adversarial Koopman models for non-linear reaction-diffusion problems that involve 
partial differential equations (PDEs). Specifically, we are interested in the (1) Kuramoto-Sivashinsky (KS) equation \cite{Kuramoto1978, Sivashinsky1977} that is
widely used to model chaos, and the (2) Gray-Scott (GS) model \cite{GrayScott, Pearson1993} for Turing instabilities. 
These equations are first solved using well established finite difference schemes to obtain the data corpus, which is 
used to train the different Koopman models in this investigation. Note that the finite difference methods required to 
solve these equations require much finer time step $\Delta t$ and so the $\Delta t$ used to obtain the data corpus is
different from the $\Delta t$ used in the Koopman analysis (the $\Delta t$ used is summarized below).
We will use the symbol $t$ hereafter to represent the time step, not to be confused with the time instant (where the time instant is $t$ multiplied by
the respective $\Delta t$). $n_S$ = 64 is used for the KS data and $n_S$ = 32 for the GS data.

\subsubsection{Kuramoto-Sivashinsky equation}

The KS model is 1D and involves only one variable $u \left(x,t \right)$, whose dynamics is governed by the following fourth-order PDE \cite{Kuramoto1978, Sivashinsky1977}:

\begin{equation}
\frac{\partial u}{\partial t} + u \frac{\partial u}{\partial x} + \frac{\partial^2 u}{\partial x^2} + \frac{\partial^4 u}{\partial x^4} = 0
\end{equation}
in a domain $x \in$ [0, 128] with periodic boundary conditions. The KS model is chaotic and is used to model the diffusive instabilities in a laminar flame front.
The initial condition for KS is given by:

\begin{equation}
u \left(x,0 \right) = \mathrm{cos} \left( x \right) + 0.1 \,\, \mathrm{cos} \left( \frac{x}{16} \right) \,\, \left( 1 + 2\,\, \mathrm{sin} \left( \frac{x}{16} \right) \right).
\end{equation}
The system is solved using Crank-Nicholson/Adams-Bashforth (CNAB2) timestepping \cite{CNAB2} with $\Delta x = \frac{1}{8}$ and $\Delta t = \frac{1}{16}$ for 4800 time steps, with the
solution at every fourth time step saved for the data corpus. Thus, $\Delta t$ = 0.25 for the Koopman analysis and we have a total of 1200 snapshots.

\subsubsection{Gray-Scott model}

\begin{figure}[h]
\centering
\includegraphics[width=12cm]{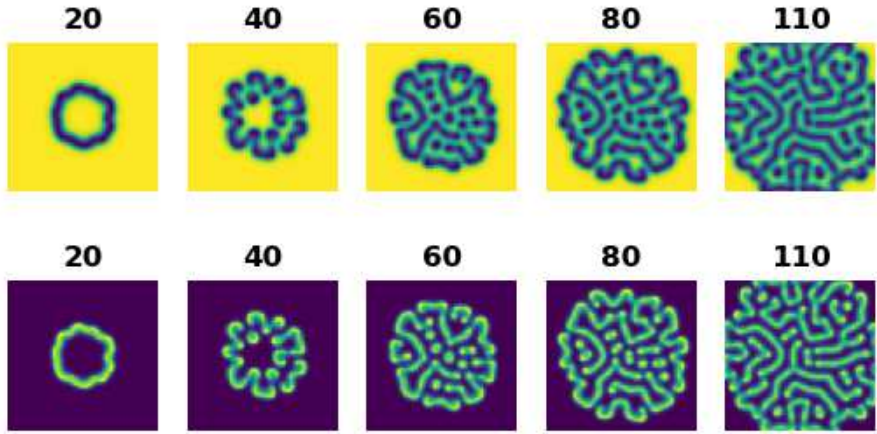}
\caption{Finite difference solutions of the Gray-Scott model at time instants 20, 40, 60, 80 and 110: $u(x,y)$ in the first row and $v(x,y)$ in the second.}
\label{fig:tur1}
\end{figure}

The GS model is 2D and involves 2 variables $u \left(x,y,t \right)$ and $v \left(x,y,t \right)$: 

\begin{eqnarray}
\frac{\partial u}{\partial t} &=& D_u \nabla^2 u - uv^2 + f \left(1-u \right) \nonumber \\
\frac{\partial v}{\partial t} &=& D_v \nabla^2 v + uv^2 - \left(f+k\right) v 
\end{eqnarray}
with $D_u$ and $D_v$ being the diffusion coefficients, $\nabla^2$ is the Laplacian operator $\left( \frac{\partial^2}{\partial x^2} + \frac{\partial^2}{\partial y^2} \right)$, and 
$f$ and $k$ are chemical reaction coefficients. Different values of these parameters gives rise to different patterns observed in the animal kingdom \cite{Pearson1993}.
The GS model is solved in a 2D box of size 2.5 $\times$ 2.5 with a mesh of size 256 $\times$ 256, with periodic boundary conditions and $\Delta t$ = 1. 
The initial condition is $\left( u,v \right)$ = $\left(1, 0 \right)$ everywhere, except in a small circular zone at the center of radius equal to
20 cells where $\left( u,v \right)$ = $\left(0.5, 0.25 \right)$ superimposed with a Gaussian noise $\mathcal{N}(\mu=0,\sigma=0.05)$. 
The values used for the transport and chemistry are: $D_u$ = 0.16, $D_v$ = 0.08, $f$ = 0.035 and $k$ = 0.060.
A second order finite difference scheme is used to solve the system for 3000 time steps and the solution at every 25-th time instant is 
saved for the data corpus. Thus, $\Delta t$ = 25 for the Koopman analysis and we have a total of 120 snapshots. Only the central 128 $\times$ 128 region of
the solution space is used for the Koopman analysis. Snapshots of the data at different time instants are presented in Fig. \ref{fig:tur1}, as evident 
one can observe Turing instability patterns \cite{Turing1952} that occur due to the coupling of species diffusion and chemical reactions. 
Turing instability is widely observed in nature, particularly in the biological world, for instance zebra stripes, giraffe polygonal patches, dalmation spots, cow patches, etc.;
more details on the physics of the Turing instability can be found in \cite{Turing1952, Pearson1993}.  

\subsection{Predictions with the deep adversarial Koopman model}
\label{sec:dakm}

\begin{figure}[h]
\centering
\includegraphics[width=12cm]{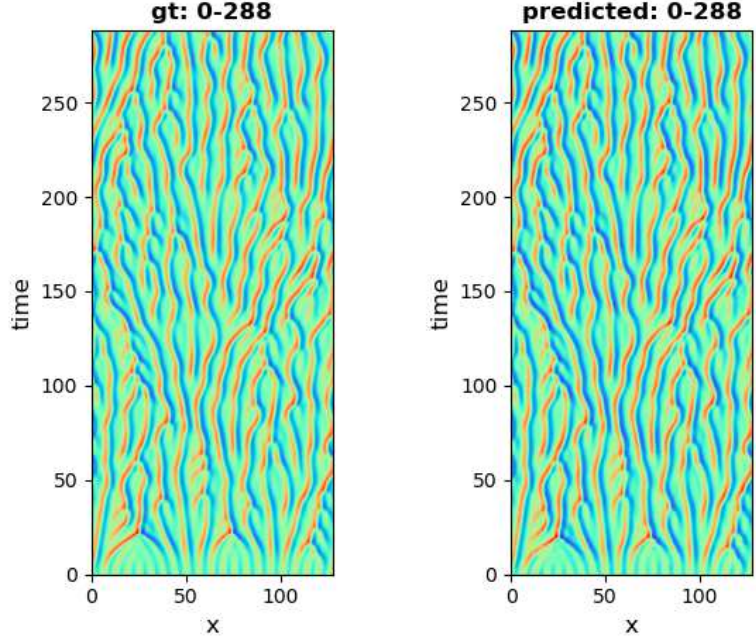}
\caption{Comparison of the ground truth (left) and prediction of the deep adversarial Koopman model (right) on the Kuramoto-Sivashinsky equation.}
\label{fig:ks1}
\end{figure}

\begin{figure}[h]
\centering
(a) \includegraphics[width=6cm]{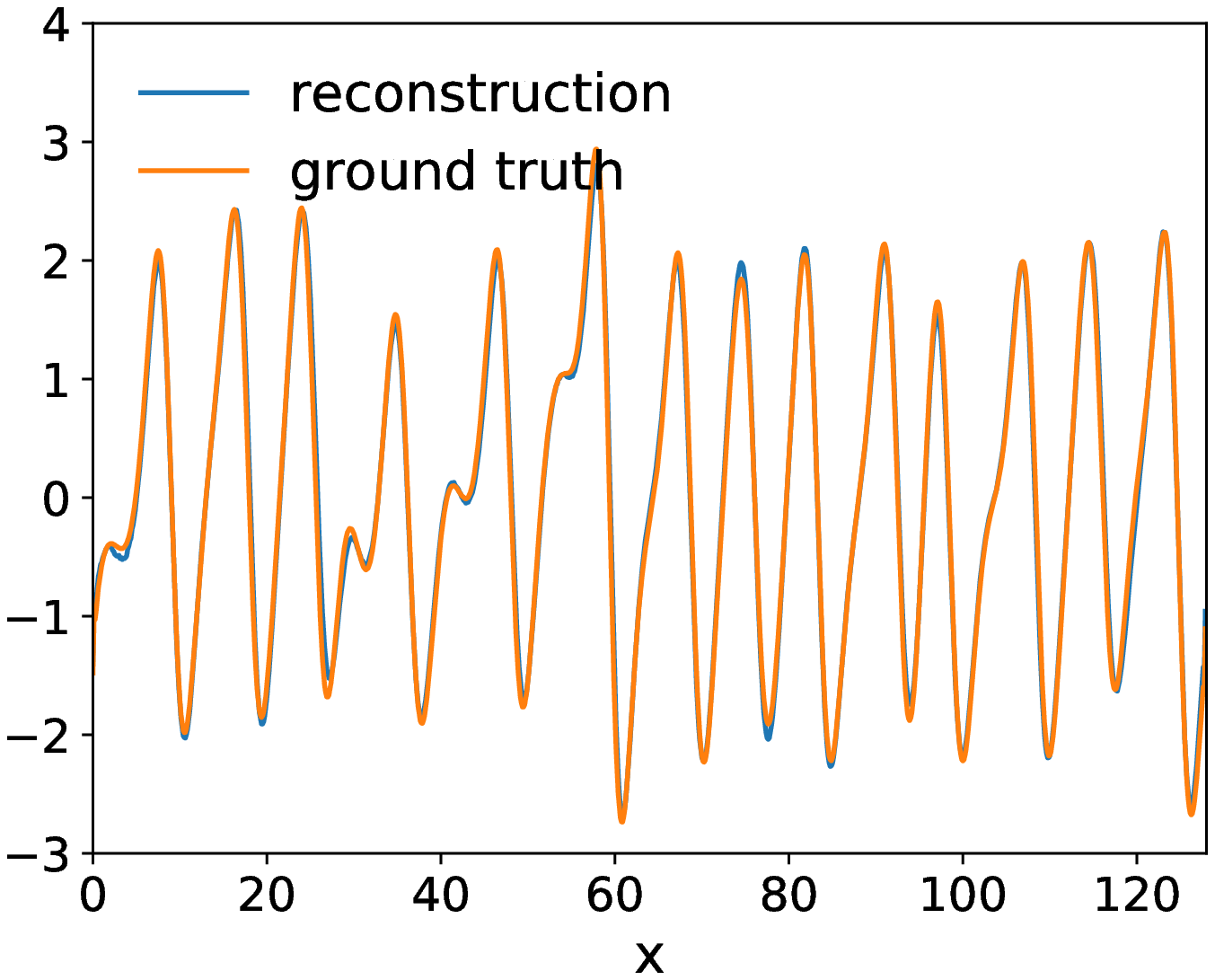}  
(b) \includegraphics[width=6cm]{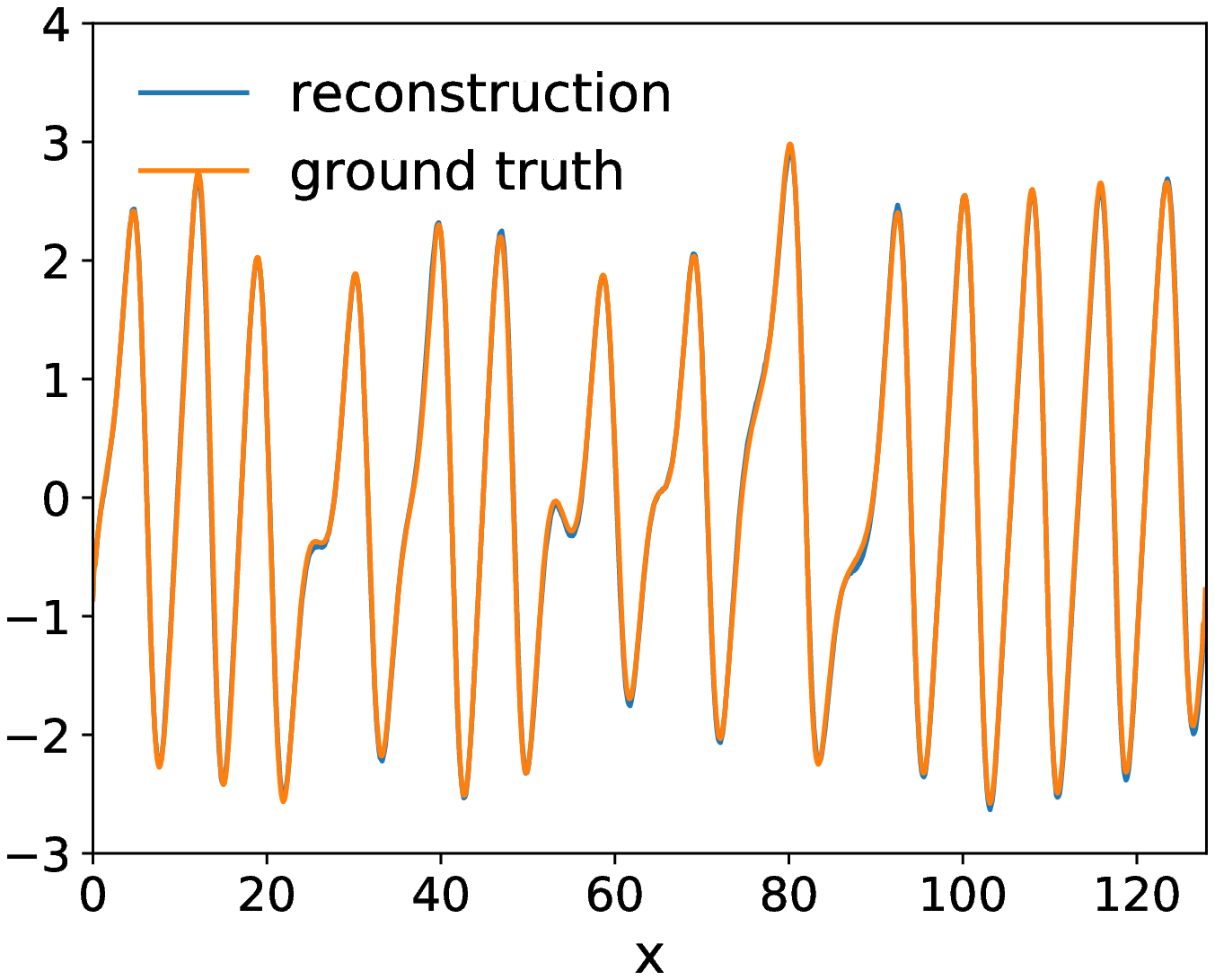}
\caption{Predictions of $u(x)$ for the Kuramoto-Sivashinsky equation with the deep adversarial Koopman model at time instants (a) 64.5 and (b) 212.5.}
\label{fig:ks2}
\end{figure}

We train the deep adversarial Koopman model on the KS data corpus using $n_S$ = 64 time-step sequences and use it to make predictions starting from time = 0 to time = 288 (i.e., for a total of 
1152 time steps, since $\Delta t$ = 0.25). This corresponds to 18 cycles, where each cycle is a sequence of 64 time steps, the last prediction from each 
cycle is used as input to the next cycle. The ground truth (gt) and the model predictions of the chaotic patterns are shown in Fig. \ref{fig:ks1}, and demonstrate good
agreement. Thus, the deep adversarial Koopman model has learned to replicate the training data. Note that this problem is not periodic and therefore it was found to 
not make accurate predictions of future states of the system, unlike the studies of \cite{Lusch2018, Morton2018} where a similar model was applied to
periodic problems such as the von Karman vortex shedding problem \cite{vonKarman}. However, our model is still able to capture the chaotic 
dynamics of the system and is able to reproduce the training data well. We show the quantitative solution at two time instants 64.5 ($t$ = 258) and 212.5 ($t$ = 850) in 
Fig. \ref{fig:ks2}, and they agree well with the ground truth finite difference solution.

\begin{figure}[h]
\centering
(a) \includegraphics[width=12cm]{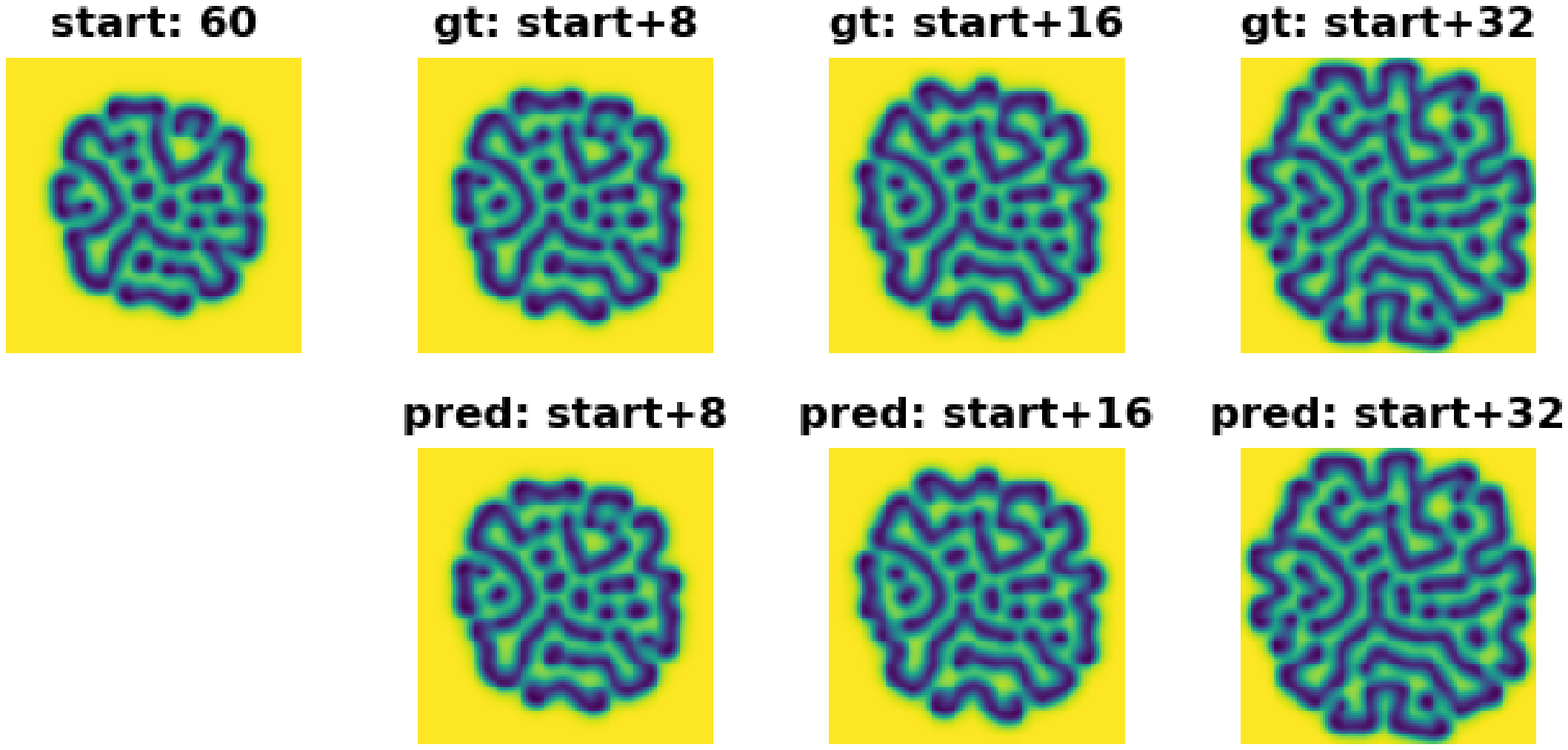} \\ 
(b) \includegraphics[width=12cm]{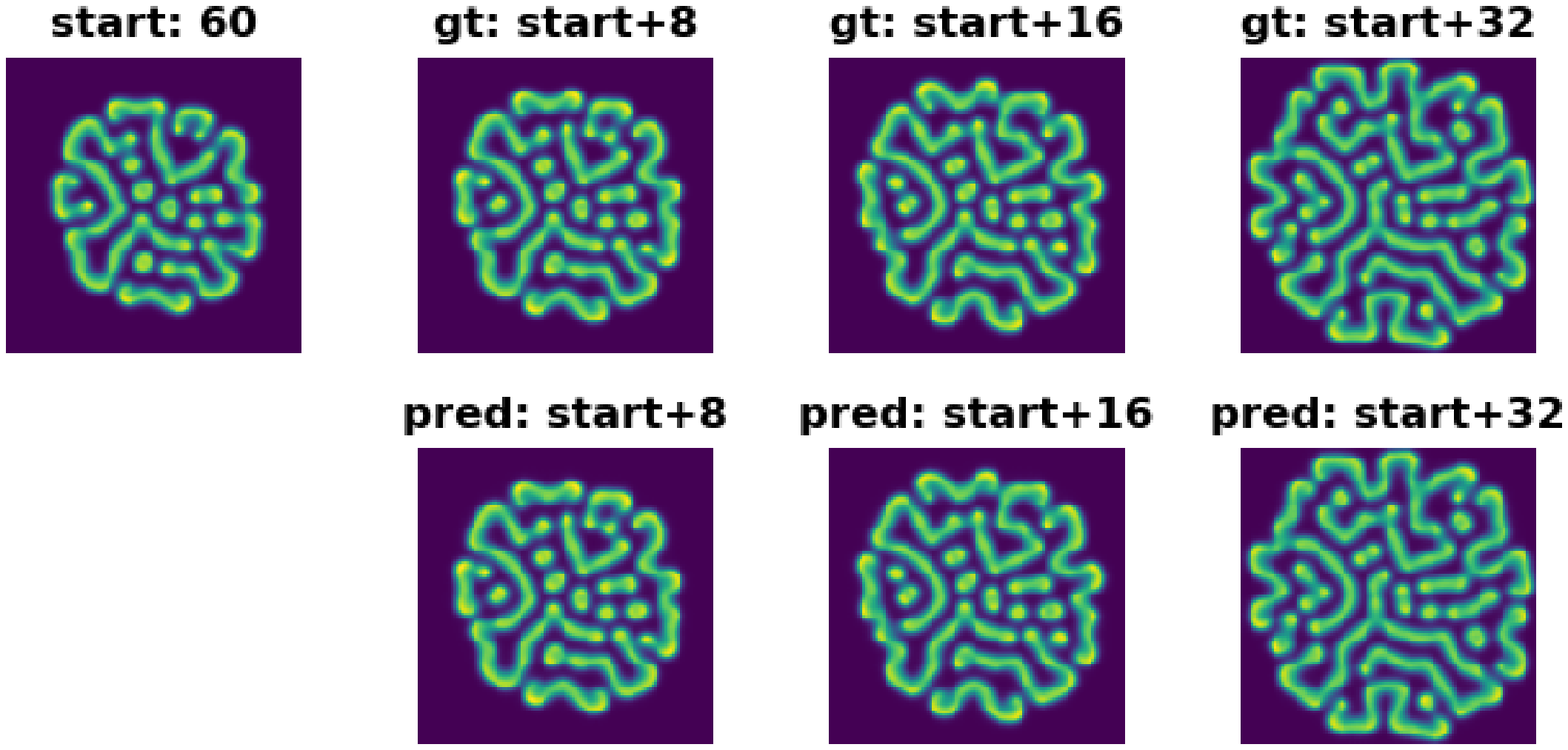}
\caption{Predictions of (a) $u$ and (b) $v$ for the Gray-Scott model starting from the solution at the 60-th time step. The top row corresponds to the ground truth (gt) Turing patterns and the model predictions are shown in the bottom row.}
\label{fig:gs1}
\end{figure}

For the GS problem, we train the deep adversarial Koopman model on the data using $n_S$ = 32 time-step sequences. 
Starting from the 60-th time step, we make predictions of the next 32 time steps, and the results for a few later
time instants are shown in Fig. \ref{fig:gs1} along with the ground truth (gt) values. As before, the model is able to replicate 
the data accurately despite the patterns not being organized in any symmetric formation. Specifically, the length scale of the 
structures as well as the diameter of the outward propagating front are accurately predicted by the model.

\subsection{Ablation study and missing data problem}

We undertake an ablation study to ascertain the preponderance of the different loss functions used in the 
deep adversarial Koopman model. In particular, we will turn off the gradient and/or GAN losses and re-train the model
to see how these terms affect the final predictions. In addition, we will also test the performance of our model on
data which has a few missing snapshots at training, which will demonstrate the model's robustness. 
For training with the missing entries, the loss terms are masked for data involving the missing entries, and so the model is tasked to 
learn the dynamics from the available entries, and use it to fill-up the missing entries. Note that since we are training long sequences of
32 or 64 time steps, the model has to learn to predict the missing entries accurately in order to be able to match the solution at the forthcoming time instants. 
In other words, suppose the solution $x_k$ at time instant $k$ is missing, with $x_{k-1}$ and $x_{k+1}$ available, the model must first learn to encode the 
available data to $z_{k-1}$ and $z_{k+1}$. It must then learn to use $z_{k-1}$ to predict the the code for the missing time step $z_{k}^{\mathrm{pred}}$, and subsequently 
use $z_{k}^{\mathrm{pred}}$ to predict the code for the following time step $z_{k+1}^{\mathrm{pred}}$. The model's prediction will be accurate if and only if
$z_{k+1}^{\mathrm{pred}} \approx z_{k+1}$. Note that the KS and the GS problems are very different in that the former is 1D and also involves fourth-order 
derivatives, whereas the latter is 2D and involves diffusion terms which are second order derivatives (i.e., Laplacian). Since the two
problems have different dynamics, our analysis takes separate approaches. For the KS data, 
we consider a joint analysis of ablation study with the missing data included; for the GS problem, the ablation and missing data 
studies are separately analyzed.

\begin{figure}[h]
\centering
\includegraphics[width=8cm]{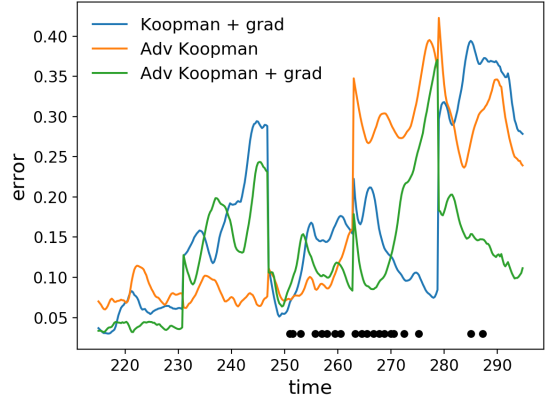}
\caption{Ablation study of the model trained using the KS problem data. Three different models are trained: (1) ``Koopman + grad'', (2) ``Adv Koopman'', and (3) ``Adv Koopman + grad''.
See the text for full description of the models. The mean L$_{1}$ error of the prediction compared to the ground truth is shown.
The black circles represent the time instants where the data is assumed to be missing at training.}
\label{fig:ks_abl}
\end{figure}

First, we consider models trained on the KS data corpus with partial missing entries. Specifically, we assume that at training time, 10\% of the data for time instants > 250 (i.e., $t >$ 1000)
is missing by setting 10\% of randomly chosen snapshots for $t >$ 1000 to zeros. Three different models are trained: (1) Koopman model with
the gradient loss, but not the adversarial loss (``Koopman + grad''), (2) Koopman model with adversarial loss but not the gradient loss (``Adv Koopman''), and (3) the
full Koopman model with both the adversarial and gradient losses (``Adv Koopman + grad''). Once the training is complete, we evaluate the three models
by starting from the ground truth solution at the time instant = 215 (i.e., $t$ = 860) and advance the dynamics using the Koopman models 
till the time instant = 295 (i.e., $t$ = 1180). The mean L$_{1}$ error between the model's prediction and
the ground truth solution from the data corpus are presented in Fig. \ref{fig:ks_abl}. The black circles 
denote the missing data at training time.
The adversarial Koopman model with the gradient loss (``Adv Koopman + grad'') performs 
relatively better than the other models. While the adversarial Koopman (``Adv Koopman'') has the lowest error near the time instant = 240,  
after the missing data values are encountered, the error increases for both the ``Adv Koopman'' and the ``Koopman + grad'' models. 
In particular, we desire models to perform well in the vicinity of the missing entries, and the ``Adv Koopman + grad'' model performs the best 
for time instants $>$ 280. Thus, the GAN and gradient losses help the model to navigate missing entries at test time more accurately as the penalty
induced by these terms will help in learning the dynamics better.

\begin{figure}[h]
\centering
\includegraphics[width=8cm]{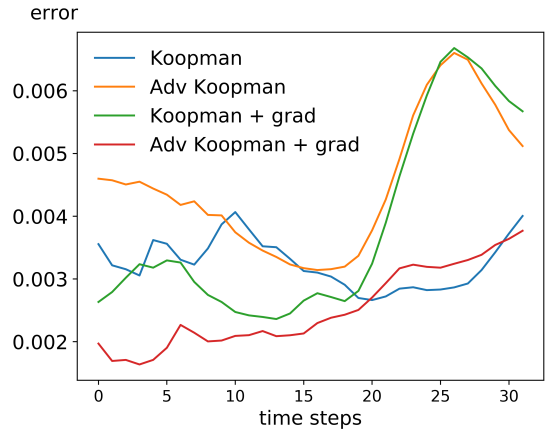} 
\caption{Ablation study of the model trained using the GS problem data. Four different models are trained on all the 120 time snapshots: (1) ``Koopman'', (2) ``Adv Koopman'', 
(3) ``Koopman + grad'', and (4) ``Adv Koopman + grad''. See the text for a full discription of the models. Shown here is the mean L$_{1}$ error of the prediction of $u(x,y,t)$ compared
to the ground truth, by starting from $t$ = 60.}
\label{fig:tur_abl}
\end{figure}

We will next demonstrate the ablation study for the Koopman models trained on the GS data corpus. 
Four variants of the model are considered: (1) Koopman model without adversarial and gradient losses (``Koopman''), (2) Koopman model with adversarial loss but not the
gradient loss (``Adv Koopman''), (3) Koopman model with the gradient loss, but not the adversarial loss (``Koopman + grad''), and (4) the full Koopman model with both
the adversarial and gradient losses (``Adv Koopman + grad''). Each of these models are trained on the full GS data corpus of 120 time steps. They are then 
evaluated as before starting from the $t$ = 60-th time step and tasked to predict the solution for the next 32 time steps. The mean L$_{1}$ error of the 
prediction of $u(x,y,t)$ compared to the ground truth is presented in Fig. \ref{fig:tur_abl}. The adversarial Koopman model with the 
gradient losses (``Adv Koopman + grad'') has the overall best performance in terms of the least error. We note that having only one of the GAN or gradient loss
seems to result in high error at later time steps, see ``Adv Koopman'' and ``Koopman + grad'' in Fig. \ref{fig:tur_abl} near 25-30 time steps. This
error is also higher than the vanilla Koopman model without these losses (``Koopman'') at the later time steps. However, when both these losses are 
jointly accounted for, the model seems to perform well (``Adv Koopman + grad'').
We conjecture that these two losses (GAN and gradient) are closely coupled in that the positive effect of one of the losses on the model 
training also influences the effect of the other. This joint coupling of the losses needs further investigation in future studies 
to ascertain if this conclusion holds for other problems of interest.

\begin{figure}[h]
\centering
(a) \includegraphics[width=3.5cm]{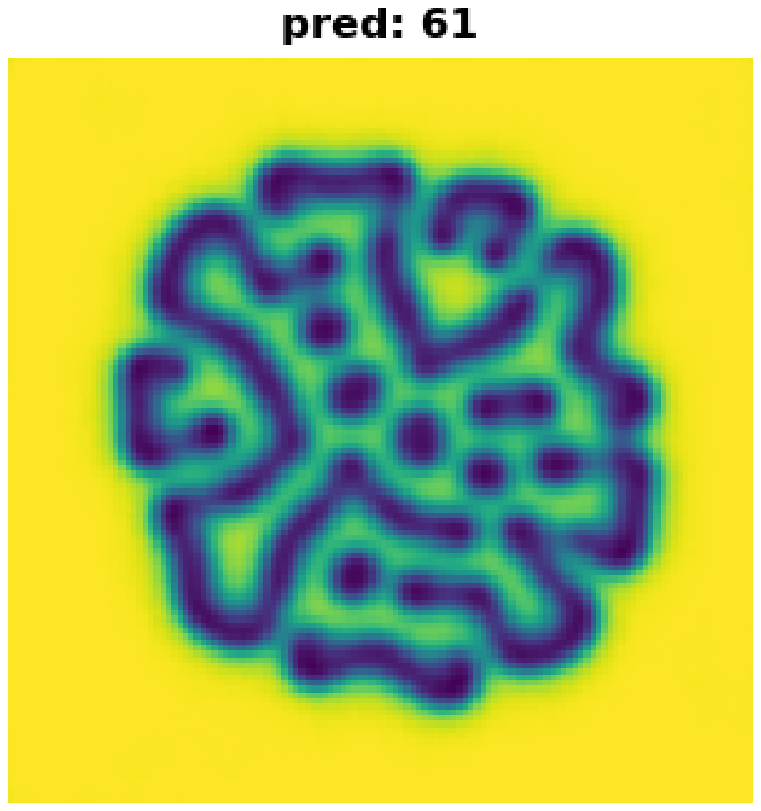} 
(b) \includegraphics[width=3.5cm]{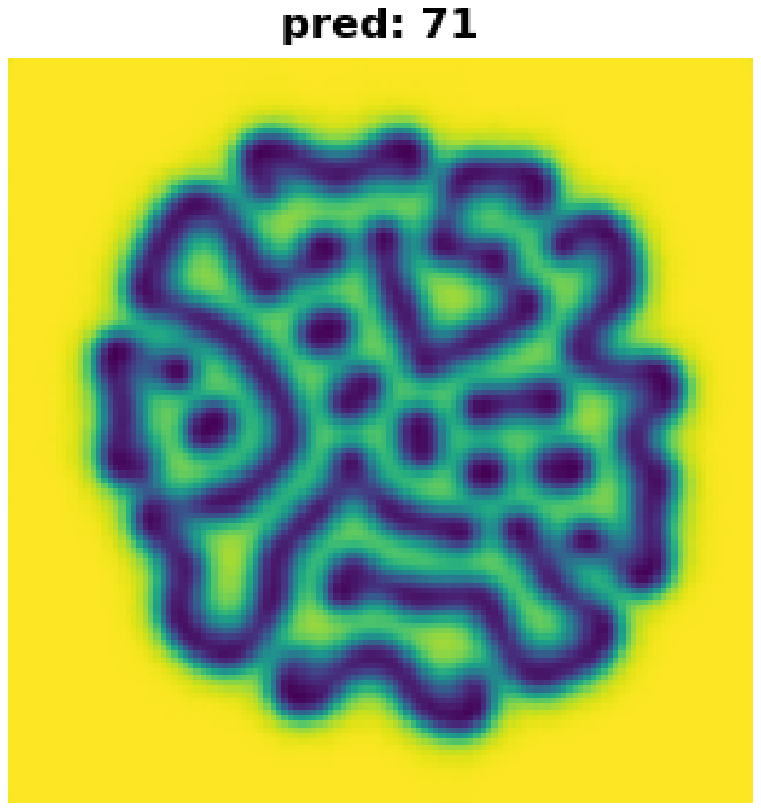}  
\caption{Missing data problem for the GS data corpus. The missing entries at $t$ = 61 and 71 as predicted by the model are shown.}
\label{fig:gs_miss}
\end{figure}

We will now consider the missing data problem for the GS data set. 5\% of the snapshots are randomly chosen and set to zero.
The adversarial Koopman model with gradient losses is trained to learn the dynamics of the system, with masks
applied to the loss function whenever a missing data is encountered. One random set of missing entries
considered are the time steps $t$ = [36, 50, 61, 71, 87, 102] (recall that for the GS problem we have 120 snapshots in time).
The model is trained on the available data and tasked to predict the missing entries, where each missing entry is predicted 
from the most recent avaialble snapshot (e.g., missing data at $t$ = 61 is predicted starting from the available data at $t$ = 60, etc.)
In Fig. \ref{fig:gs_miss}, we present the model's prediction of two missing snapshots, $t$ = 61 and 71. Note that for this 
analysis, the missing entries are unavailable at training, but the model must learn the full dynamics of time evolution and 
use that to predict the missing values. The predicted Turing patterns observed in Fig. \ref{fig:gs_miss} are reasonable, and the 
errors with respect to the ground truth values at these time instants are also low (not shown). Thus, the model is 
able to predict missing entries in data.

\section{Control for forced system evolution}

The inherent linearization of the dynamics in the Koopman model allows us to also undertake a 
control approach akin to classical linearized control, for which there are copious algorithms. 
A similar control analysis using the Koopman model was also undertaekn in \cite{Morton2018}
where the von Karman vortices behind a cylinder were suppressed by adding control inputs to the 
system dynamics. Specifically, the linearized control dynamics can be summarized as (note the residual $z_t$ component):

\begin{equation}
z_{t+1} = z_t + K z_t + L U_t,
\label{eqn:control1}
\end{equation}
where $L$ is a prescribed control matrix and $U$ is a set of control inputs that can be fed into
the system to achieve a desired future system dynamics. Note that $z \in \mathcal{R}^M$, $K \in \mathcal{R}^{M \times M}$,
$L \in \mathcal{R}^{M \times M}$, and $U_t \in \mathcal{R}^{M}$.
We take a similar approach, but here the goal of our control analysis is to
provide a control matrix, which the model must use to learn control inputs $U_t$ so that 
the future course of the system dynamics can be altered as desired. 
In this setting, we fix $L$ and use gradient descent to learn $U_t$ for a desired objective. 
The GS data corpus is used for this control analysis.

\begin{figure}[h]
\centering
(a) \includegraphics[width=8cm]{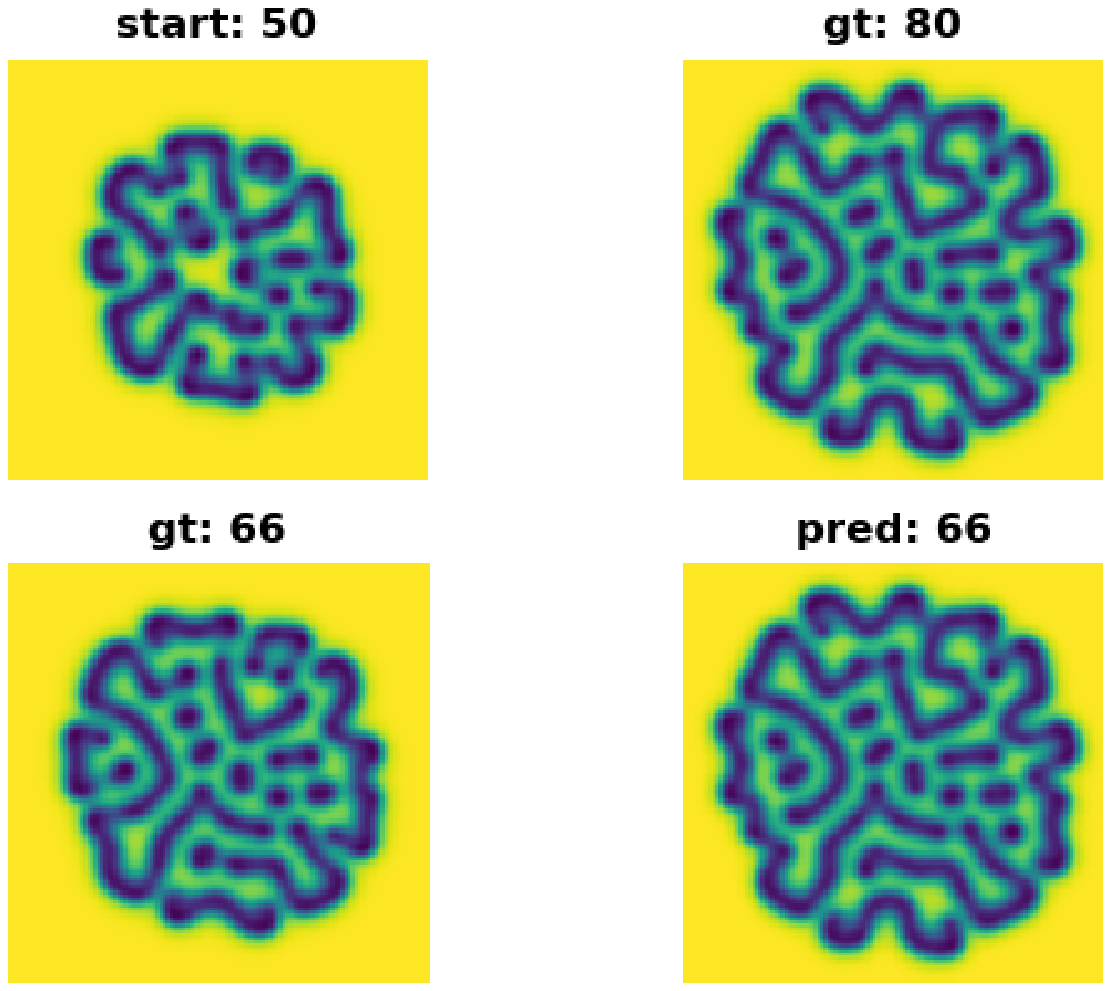} 
(b) \includegraphics[width=4cm]{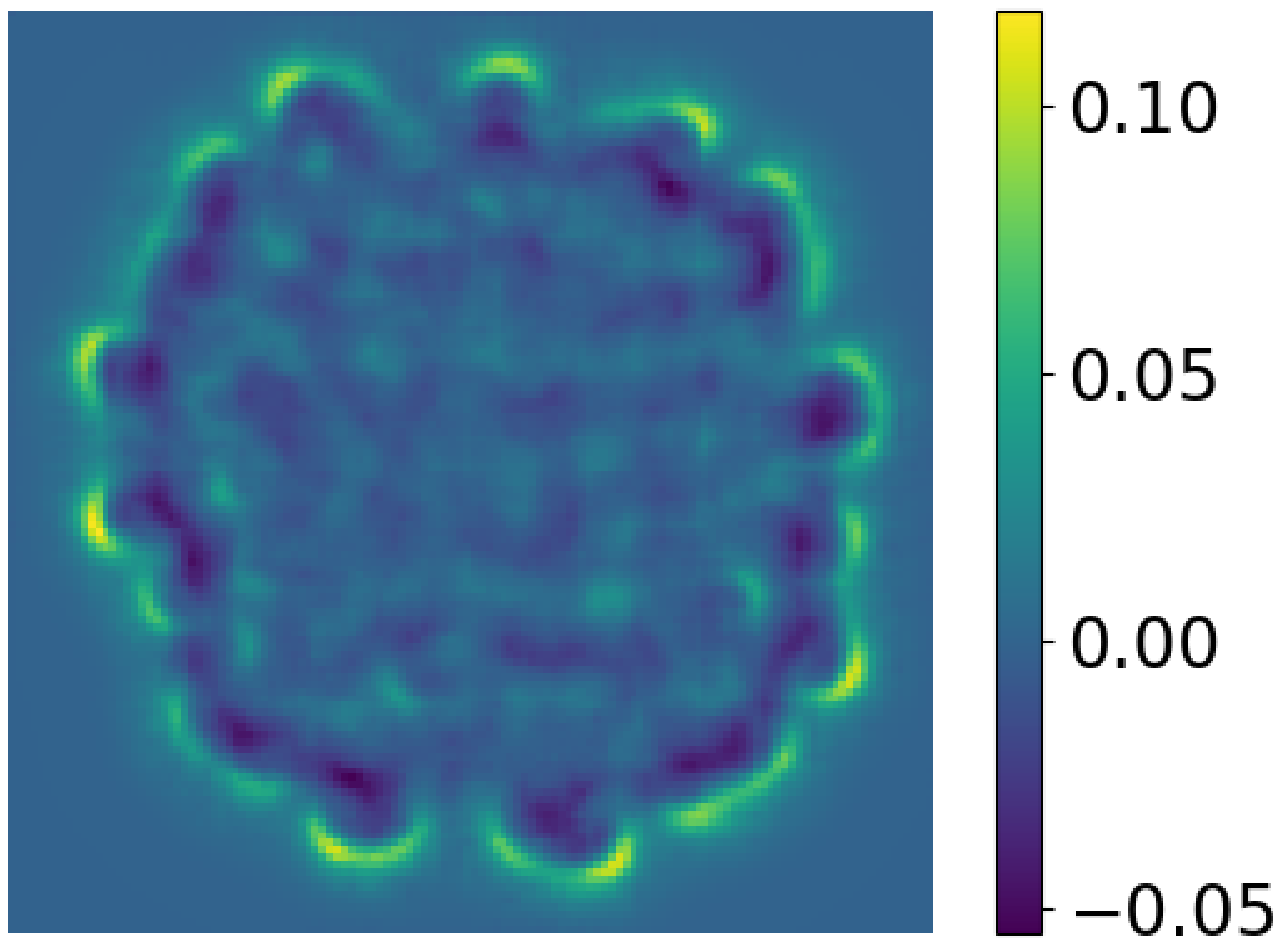} 
\caption{Control for forcing the system evolution. The start ($t$=50), desired (same as ground truth at $t$ = 80), predicted ($t$ = 50+16) and ground truth at $t$ = 66 flow-field for $u(x,y,t)$ are shown in (a). The error between the predicted and the desired is shown in (b).}
\label{fig:tur_control1}
\end{figure}

The model is tasked with the acceleration of the system dynamics using control, so that the 
Turing instability structures evolve faster than their natural rate.  
Specifically, we consider the ground truth solution at time step $t = t_\mathrm{start}$, and learn 
control inputs $U_t$ for a fixed $L$ matrix so that the prediction of the flow field after $\delta$ time
steps matches that at $t = t_\mathrm{desired}$. That is, starting from the ground truth flow field at $t_\mathrm{start}$, we want the
prediction of the flow field at $t_{\mathrm{start} + \delta}$ to look like the ground truth flow field at $t_\mathrm{desired}$, thereby 
evolving the system at a forced rate. For the analysis, we will use $t_\mathrm{start}$ = 50, $t_\mathrm{desired}$ = 80 and $\delta$ = 16.
$L$ is fixed to be a randomly initialized 
$M \times M$ boolean matrix with entries that are either 1 with a probability of 0.4, or 0 otherwise.  
The baseline Koopman model we use is the Adversarial Koopman model with gradient losses, trained on the full
GS corpus of 120 time snapshots as considered in Section \ref{sec:dakm}. The model weights are fixed for this control
analysis and only the $U_t$ values are learned by minimizing the following loss:

\begin{equation}
L^{\mathrm{control}} = \left( z_{\mathrm{start}+\delta}^{\mathrm{pred}} - z_{\mathrm{desired}}^{\mathrm{ground \, truth}} \right )^2 + \sum_{t=t_{\mathrm{start}}}^{t=t_{\mathrm{start}} + \delta-1} U_{t}^2.
\end{equation}
The second term in the above loss function enforces the constraint that $U_t$ values are not large. 
In Fig. \ref{fig:tur_control1}(a) we show the starting flow-field for $u(x,y,t)$ (at $t$ = 50), the desired
flow-field 16 time steps into the future (which we want to be the actual flow-field at $t$ = 80 from the data corpus), and the 
predicted flow-field of $u(x,y)$ by the model at $t$ = 66 (i.e., 16 time steps from the start). We also show
the actual solution at $t$ = 66 in the data corpus, which we are trying to circumvent by adding the control inputs. 
As evident, the model has learned the required control inputs so that the evolution of the Turing structures can be
accelerated. The error $u^{gt}(x,y,t=t_{\mathrm{desired}}) - u^{\mathrm{pred}}(x,y,t=t_{\mathrm{start}}+\delta)$ is shown in Fig. \ref{fig:tur_control1}(b). While the 
model predicts near zero errors in most of the regions of the flow-field, at the outer front of the instability, the errors are
slightly high, but are concentrated only in a thin zone at the outer edges of the Turing-instability where the 
reaction rates are high. We believe that the relatively higher errors at the outer edges of the instability is because 
with the control forcing $U_t$, the system evolves at an unnatural rate, and the ensuing latent codes $z_{t}$ are unique to the 
forced inputs, and not seen by the decoder at training, which gives rise to slightly higher errors at the 
outer edges of the Turing instability front. $v(x,y,t)$ is not shown for brevity, but similar conclusions also hold.
We repeated the problem where the decoder is also fine-tuned using gradient 
descent along with the control inputs $U_t$ and here the errors were near zero everywhere (not shown), including at the outer edges of the Turing
instability. While this is more accurate, re-training the decoder is not desired as we consider this analysis as a control-only problem. 
Future work must address how to better handle control when the $g$ and $g^{-1}$ are fixed and only the control inputs are learned.

\section{Related work}

Even though the Koopman model \cite{Koopman1931, Koopman1932} was introduced almost a century ago, 
there is renewed interest in improving it to model many dynamical systems. 
The Dynamic Mode Decomposition (DMD) algorithm \cite{Kutz2016} is one approach to 
represent the Koopman operator. Other deep learning-based approaches such as autoencoders that learn  
to encode from the physical to the Koopman invariant subspace, and vice versa in the decoder, are becoming popular \cite{Takeishi2017, Lusch2018, Morton2018}.
While the Koopman operator can be computed by solving a least squares optimization as in \cite{Morton2018},
it was evaluated from an auxiliary network and eigenvalue treatment in \cite{Lusch2018}.
The latter approach---which is extended in this study---is very promising as highly non-linear functions that are
non-periodic can also be studied using the Koopman treatment.

As the Deep Learning community develops more novel algorithms to reprsent data, some of these
can be used in lieu of autoencoders and alternate Koopman formulations can be reconstructed. 
For instance, videos that are a sequence of individual frames can be treated using a Koopman approach to build a 
model of the dynamics. Even the field of model-based Reinforcement Learning \cite{Dyna, Nagabandi2017} can use a Koopman-type 
model for learning the system dynamics $p(x_{t+1}|x_t,a_t)$. Here, the state of the system $x_t$ can be very high dimensional    
and so one can use a deep learning approach to embed the state to a Koopman invariant sub-space, where the action $a_t$ will advance
the model dynamics, which can be decoded to the next state $x_{t+1}$ of the system. 
The fluid dynamics community can also greatly benefit with the use of deep learning-based Koopman models, as also
evidenced by many recent publications \cite{Lusch2018, Morton2018, Pan2020}. 
It suffices to say that the applications of deep learning-based Koopman models are plentiful.


\section{Conclusion}

We have developed and presented a method based on Koopman family of algorithms to train dynamical systems.
Our model couples a GAN discriminator with a Deep Koopman model, with an auxiliary
network used to obtain the Koopman matrix, $K$. The model is robust at learning the 
dynamics of two varieties of reaction-diffusion systems: (1) the Kuramoto-Sivashinsky equation 
for chaos and (2) Gray-Scott model for Turing instabilities. Ablation studies are also conducted 
to demonstrate the efficacy of the model presented in this paper.
We have also extended the model to handle missing data, where the model is tasked to learn the 
dynamics in the presence of a few missing snapshots, and then predict these missing entries
from the learned dynamics. Furthermore, we have also extended the model to handle control
inputs in the Koopman invariant subspace and have demonstrated how one can use this to accelerate the 
growth of the Turing instabilities in time. In the future, we hope to apply the proposed 
method to other flow problems, as well as in model-based Reinforcement Learning to learn the system dynamics.   


\section*{Appendix A: neural network architectures}

The neural network architectures are summarized here. 
Note that we have a total of 4 neural networks: \textit{Encoder}, \textit{Decoder}, auxiliary network \textit{AUX} (to obtain $K$) and the GAN discriminator \textit{DISC}.  
We will use several different deep learning building blocks: batch normalization (\textit{BN}) \cite{BN}, Dropout (\textit{Dropout}) \cite{dropout},
convolutional (\textit{conv}) and deconvolutional (\textit{dconv}) \cite{dconv} operators, and the Relu (\textit{Relu}) activation function.
We will use the notation \textit{conv}(k,f,S,s) for a convolutional layer with kernel size $k$, $f$ filters, same padding (identified by $S$) and a stride of $s$. 
In addition, we will use the notation \textit{Dense}($n$) to refer to a fully connected layer with $n$ neurons. 
We first define a bottleneck layer for \textit{Encoder} with $N_f$ filters as input, \textit{BottleNeck}$^{e}$($N_f$), comprising of the following in the same order:    
\textit{BottleNeck}$^{e}$($N_f$) = \textit{BN} $\rightarrow$ \textit{Relu} $\rightarrow$  \textit{conv}(1,$N_f$/2,S,1) $\rightarrow$  \textit{BN} $\rightarrow$  \textit{Relu} $\rightarrow$  \textit{conv}(3,$N_f$/2,S,1) $\rightarrow$  \textit{BN} $\rightarrow$  \textit{Relu} $\rightarrow$  \textit{conv}(1,$N_f$,S,1).
Stated in these terms, \textit{Encoder} consists of 5 layers of convolutional operations supplemented with bottleneck layers added residually, similar to Resnet \cite{Resnet}. 
For ease of notation, we will refer to \textit{Encoder}'s residual block as \textit{RES}$^{e}$($N_f$) = \textit{conv}(3,$N_f$,S,2) + \textit{BottleNeck}$^{e}$($N_f$).
Specifically, \textit{Encoder} consists of 5 residual layers in succession: \textit{RES}$^{e}$(64) $\rightarrow$ \textit{RES}$^{e}$(128) $\rightarrow$ \textit{RES}$^{e}$(256) $\rightarrow$
\textit{RES}$^{e}$(512) $\rightarrow$ \textit{RES}$^{e}$(512), followed by a \textit{Relu} and a flattening operation. This is then fed into a \textit{Dense}($M$) layer, where $M$ 
is the dimension of the encoded embedding in the Koopman invariant subspace (we will use $M$ = 64 throughout this study). The final output of  
\textit{Encoder} does not go through any activation function.     

For the \textit{Decoder} we define a similar bottleneck layer with $N_f$ filters, albeit this time using deconvolutional operations:  
\textit{BottleNeck}$^{d}$($N_f$) = \textit{BN} $\rightarrow$ \textit{Relu} $\rightarrow$ \textit{dconv}(1,$N_f$/2,S,1) $\rightarrow$ \textit{BN} $\rightarrow$ \textit{Relu} $\rightarrow$ \textit{dconv}(3,$N_f$/2,S,1) $\rightarrow$ \textit{BN} $\rightarrow$ \textit{Relu} $\rightarrow$ \textit{dconv}(1,$N_f$,S,1). Furthermore, the 
residual block for \textit{Decoder} is different from that used for \textit{Encoder}. For \textit{Decoder}, we first add the input to the bottleneck layer akin to
Resnet \cite{Resnet}, which is then passed though a deconvolutional layer, like so: \textit{RES}$^{d}$($N_f$) = input + \textit{BottleNeck}$^{d}$($N_f$) $\rightarrow$ \textit{dconv}(3,$N_f$,S,2).
\textit{Decoder} starts with \textit{Dense}($\cdot$) with the number of neurons used being the same as the dimension of the \textit{Encoder}'s flattened output. 
This is reshaped as appropriate and is followed by 5 layers of the \textit{Decoder}'s residual blocks in succession: \textit{RES}$^{d}$(512) $\rightarrow$ \textit{RES}$^{d}$(256) $\rightarrow$ \textit{RES}$^{d}$(128) $\rightarrow$ \textit{RES}$^{d}$(64) $\rightarrow$ \textit{RES}$^{d}$($n_{out}$). Here, $n_{out}$ is the number of output channels in the data, with $n_{out}$ = 1 for the KS data corpus, and $n_{out}$ = 2 for the GS problem. For the GS problem, the output of \textit{Decoder} represents concentration that is bounded in the $\left[0,1\right]$ range, and so the sigmoid activation function is used at the end. For the KS problem, the output represents the variable $u(x,y,t)$, which is not bounded and so no activation function is used at \textit{Decoder}'s output layer.   
The $\textit{conv}$ and $\textit{dconv}$ are 1D operations for the KS problem and 2D for GS.

For the \textit{AUX} network, we define a fully connected layer with $N$ neurons as \textit{FC}($N$) = \textit{Dense}($N$) $\rightarrow$ \textit{Relu} $\rightarrow$ \textit{Dropout}. 
For $\textit{Dropout}$, we set the probability of keeping the activations to 0.8 at training, and 1.0 at testing.   
\textit{AUX} network consists of 4 fully connected layers: \textit{FC}(128) $\rightarrow$ \textit{FC}(256) $\rightarrow$ \textit{FC}(512) $\rightarrow$ \textit{Dense}($M^2$). 
The final output is then reshaped as a $M \times M$ matrix and represents the Koopman matrix, $K$.   

For \textit{DISC}, we will use the Leaky Relu activation function, denoted as \textit{LRelu}, with a slope of 0.2 in the negative side.
We define a block \textit{B}$^\mathrm{DISC}$($N_f$) as \textit{conv}(5,$N_f$,S,2) $\rightarrow$ \textit{BN} $\rightarrow$ \textit{LRelu}.
\textit{DISC} is then constructed as: \textit{conv}(5,64,S,2) $\rightarrow$ \textit{LRelu} $\rightarrow$ \textit{B}$^\mathrm{DISC}$(128) $\rightarrow$ \textit{B}$^\mathrm{DISC}$(256) $\rightarrow$ \textit{B}$^\mathrm{DISC}$(512). The output is then reshaped and passed to a \textit{Dense}(1) without any activation function to represent the Wasserstein distance. 
 
Adam \cite{Adam} optimizer is used to train the neural networks with a learning rate of 5$\times$10$^{-5}$.
A total of 50000 iterations are considered for each training, where at each iteration step one sequence
of $n_{S}$ contiguous snapshots are randomly sampled from the data corpus and used to train the networks. For the KS problem, $n_{S}$ = 64, and for the 
GS problem, $n_{S}$ = 32.  

\section*{Appendix B: gradient losses}

As aforementioned, we consider two reaction-diffusion systems: the Kuramoto-Sivashinsky equation and the Gray-Scott model, and these
are referred to as KS and GS, respectively. 
The KS problem involves first, second and fourth derivatives in the governing equation, and so we consider all the three gradient losses, with
finite difference approximations used for evaluating them. For the KS problem, we use $\lambda_1$ = 1, $\lambda_2$ = 10$^{-5}$ and $\lambda_4$ = 10$^{-8}$.
For the GS problem, we used $\lambda_1$ = 1, $\lambda_2$ = $\lambda_4$ = 0. 
The finite difference approximations for any variable $u$ in 1D are:

\begin{eqnarray}
\frac{\partial u}{\partial x} \approx \frac{u_{i+1} - u_{i-1}}{2\Delta x}, \nonumber \\
\frac{\partial^2 u}{\partial x^2} \approx \frac{u_{i+1} - 2 u_{i} + u_{i-1}}{\Delta x^2}, \nonumber \\
\frac{\partial^4 u}{\partial x^4} \approx \frac{u_{i+2} - 4 u_{i+1} + 6 u_{i} - 4 u_{i-1} + u_{i-2}}{\Delta x^4},
\end{eqnarray}
where $i$ represents the discretization index.
Extension to 2D is straightforward and not presented here for brevity.


\bibliographystyle{unsrt}
\bibliography{DeepAdvKoopman}

\begin{thebibliography}{10}

\bibitem{POD1}
K.~Kunisch and S.~Volkwein.
\newblock Galerkin proper orthogonal decomposition methods for a general
  equation in fluid dynamics.
\newblock {\em SIAM Journal on Numerical Analysis}, 40(2):492--515, 2002.

\bibitem{POD2}
J.~Burkardt, M.~Gunzburger, and H.~C. Lee.
\newblock Pod and cvt-based reduced-order modeling of navier–stokes flows.
\newblock {\em Computer Methods in Applied Mechanics and Engineering},
  196:337--355, 2006.

\bibitem{DEIM}
S.~Chaturantabut and D.~C. Sorensen.
\newblock Nonlinear model reduction via discrete empirical interpolation.
\newblock {\em SIAM Journal on Scientific Computing}, 32(5):2737--2764, 2010.

\bibitem{Koopman1931}
B.~O. Koopman.
\newblock Hamiltonian systems and transformation in hilbert space.
\newblock {\em Proceedings of the National Academy of Sciences USA},
  17:315--318, 1931.

\bibitem{Koopman1932}
B.~O. Koopman and J.~von Neumann.
\newblock Dynamical systems of continuous spectra.
\newblock {\em Proceedings of the National Academy of Sciences USA},
  18:255--263, 1932.

\bibitem{Schmid2010}
P.~J. Schmid.
\newblock Dynamic mode decomposition of numerical and experimental data.
\newblock {\em Journal of Fluid Mechanics}, 656:5--28, 2010.

\bibitem{Kutz2016}
J.~N. Kutz, S.~L. Brunton, B.~W. Brunton, and J.~L. Proctor.
\newblock Dynamic mode decomposition: Data-driven modeling of complex systems.
\newblock {\em Society for Industrial and Applied Mathematics}, 2016.

\bibitem{AE}
G.~E. Hinton and R.~R. Salakhutdinov.
\newblock Reducing the dimensionality of data with neural networks.
\newblock {\em Science}, 313:504--507, 7 2006.

\bibitem{Backprop}
D.~E. Rumelhart, G.~E. Hinton, and R.~J. Williams.
\newblock Learning representations by back-propagating errors.
\newblock {\em Nature}, 323:533--536, 1986.

\bibitem{Hornik1989}
K.~Hornik, M.~Stinchcombe, and H.~White.
\newblock Multilayer feedforward networks are universal approximators.
\newblock {\em Neural Networks}, 2(5):359--366, 1989.

\bibitem{Hornik1991}
K.~Hornik.
\newblock Approximation capabilities of multilayer feedforward networks.
\newblock {\em Neural Networks}, 4(2):251--257, 1991.

\bibitem{GAN}
I.~J. Goodfellow, J.~Pouget-Abadie, M.~Mirza, B.~Xu, D.~Warde-Farley, S.~Ozair,
  A.~Courville, and Y.~Bengio.
\newblock Generative adversarial nets.
\newblock {\em Advances in Neural Information Processing Systems (NIPS)}, 2014.

\bibitem{Takeishi2017}
N.~Takeishi, Y.~Kawahara, and T.~Yairi.
\newblock Learning koopman invariant subspaces for dynamic mode decomposition.
\newblock {\em Advances in Neural Information Processing Systems (NIPS)}, 2017.

\bibitem{Yeung2017}
E.~Yeung, S.~Kundu, and N.~Hodas.
\newblock Learning deep neural network representations for koopman operators of
  nonlinear dynamical systems.
\newblock {\em Arxiv:1708.06850}, 2017.

\bibitem{Lusch2018}
B.~Lusch, J.~N. Kutz, and S.~L. Brunton.
\newblock Deep learning for universal linear embeddings of nonlinear dynamics.
\newblock {\em Nature Communications}, 9(1):4950, 2018.

\bibitem{Morton2018}
J.~Morton, F.~D. Witherden, A.~Jameson, and M.~J. Kochenderfer.
\newblock Deep dynamical modeling and control of unsteady fluid flows.
\newblock {\em Advances in Neural Information Processing Systems (NIPS)}, 2018.

\bibitem{Morton2019}
J.~Morton, F.~D. Witherden, and M.~J. Kochenderfer.
\newblock Deep variational koopman models: inferring koopman observations for
  uncertainty-aware dynamics modeling and control.
\newblock {\em Arxiv: 1902.09742}, 2019.

\bibitem{vonKarman}
T.~von Karman.
\newblock Aerodynamics.
\newblock {\em McGraw-Hill}, 1963.

\bibitem{Kuramoto1978}
Y.~Kuramoto.
\newblock Diffusion-induced chaos in reaction systems.
\newblock {\em Progress of Theoretical Physics Supplement}, 64:346--367, 1978.

\bibitem{Sivashinsky1977}
G.~I. Sivashinsky.
\newblock Nonlinear analysis of hydrodynamic instability in laminar flames-i.
  derivation of basic equations.
\newblock {\em Acta Astronautica}, 4(11-12):1177--1206, 1977.

\bibitem{GrayScott}
P.~Gray and S.~K. Scott.
\newblock Autocatalytic reactions in the isothermal, continuous stirred tank
  reactor: Oscillations and instabilities in the system a + 2b $\rightarrow$
  3b; b $\rightarrow$ c.
\newblock {\em Chemical Engineering Science}, 39(6):1087--1097, 1984.

\bibitem{Pearson1993}
J.~E. Pearson.
\newblock Complex patterns in a simple system.
\newblock {\em Science}, 261(5118):189--192, 1993.

\bibitem{Proctor2014}
J.~L. Proctor, S.~L. Brunton, and J.~N. Kutz.
\newblock Dynamic mode decomposition with control.
\newblock {\em Arxiv: 1409.6358}, 2014.

\bibitem{Kaiser2017}
E.~Kaiser, J.~N. Kutz, and S.~L. Brunton.
\newblock Data-driven discovery of koopman eigenfunctions for control.
\newblock {\em Arxiv: 1707.01146}, 2017.

\bibitem{Mezic2004}
I.~Mezic and A.~Banaszuk.
\newblock Comparison of systems with complex behavior.
\newblock {\em Physica D}, 197:101--133, 2004.

\bibitem{Mezic2013}
I.~Mezic.
\newblock Analysis of fluid flows via spectral properties of the koopman
  operator.
\newblock {\em Annual Review of Fluid Mechanics}, 45:357--378, 2013.

\bibitem{Roshko1955}
A.~Roshko.
\newblock On the wake and drag of bluff bodies.
\newblock {\em Journal of the Aeronautical Sciences}, 22(2):124--132, 1955.

\bibitem{VAEGAN}
A.~B.~L. Larsen, S.~K. Sonderby, H.~Larochelle, and O.~Winther.
\newblock Autoencoding beyond pixels using a learned similarity metric.
\newblock {\em Arxiv: 1512.09300}, 2016.

\bibitem{WGANGP}
I.~Gulrajani, F.~Ahmed, M.~Arjovsky, V.~Dumoulin, and A.~C. Courville.
\newblock Improved training of wasserstein gans.
\newblock {\em Advances in Neural Information Processing Systems (NIPS)}, 2017.

\bibitem{Mathieu2015}
M.~Mathieu, C.~Couprie, and Y.~LeCun.
\newblock Deep multi-scale video prediction beyond mean square error.
\newblock {\em Arxiv: 1511.05440}, 2015.

\bibitem{CNAB2}
D.~Wang and S.~J. Ruuth.
\newblock Variable step-size implicit-explicit linear multistep methods for
  time-dependent partial differential equations.
\newblock {\em Journal of Computational Mathematics}, 26(6):838--855, 2008.

\bibitem{Turing1952}
A.~Turing.
\newblock The chemical basis of morphogenesis.
\newblock {\em Philosophical Transactions of the Royal Society of London B},
  237(641):37--72, 1952.

\bibitem{Dyna}
R.~S. Sutton.
\newblock Dyna, an integrated architecture for learning, planning, and
  reacting.
\newblock {\em SIGART Bulletin}, 2:160--163, 1991.

\bibitem{Nagabandi2017}
A.~Nagabandi, G.~Kahn, R.~S. Fearing, and S.~Levine.
\newblock Neural network dynamics for model-based deep reinforcement learning
  with model-free fine-tuning.
\newblock {\em Arxiv:1708.02596}, 2017.

\bibitem{Pan2020}
S.~Pan and K.~Duraisamy.
\newblock Physics-informed probabilistic learning of linear embeddings of
  nonlinear dynamics with guaranteed stability.
\newblock {\em SIAM Journal on Applied Dynamical Systems}, 19(1):480--509,
  2020.

\bibitem{BN}
S.~Ioffe and C.~Szegedy.
\newblock Batch normalization: accelerating deep network training by reducing
  internal covariate shift.
\newblock {\em Proceedings of the 32nd International Conference on Machine
  Learning}, 2015.

\bibitem{dropout}
N.~Srivastava, G.~Hinton, A.~Krizhevsky, I.~Sutskever, and R.~Salakhutdinov.
\newblock Dropout: A simple way to prevent neural networks from overfitting.
\newblock {\em Journal of Machine Learning Research}, 15(56):1929--1958, 2014.

\bibitem{dconv}
V.~Dumoulin and F.~Visin.
\newblock A guide to convolution arithmetic for deep learning.
\newblock {\em Arxiv:1603.07285}, 2018.

\bibitem{Resnet}
K.~He, X.~Zhang, S.~Ren, and J.~Sun.
\newblock Deep residual learning for image recognition.
\newblock {\em IEEE Conference on Computer Vision and Pattern Recognition
  (CVPR)}, 2016.

\bibitem{Adam}
D.~P. Kingma and J.~Ba.
\newblock Adam: a method for stochastic optimization.
\newblock {\em 3rd International Conference for Learning Representations
  (ICLR)}, 2015.

\end{thebibliography}

\end{document}